\documentclass[a4paper,conference]{IEEEtran}
\IEEEoverridecommandlockouts
\usepackage{cite}
\usepackage{amsmath,amssymb,amsfonts}
\usepackage{algorithmic}
\usepackage{graphicx}
\usepackage{textcomp}
\usepackage{xcolor}

\def\BibTeX{{\rm B\kern-.05em{\sc i\kern-.025em b}\kern-.08em
    T\kern-.1667em\lower.7ex\hbox{E}\kern-.125emX}}

\usepackage{subfigure}
\usepackage[breaklinks]{hyperref}
\usepackage{breakurl}

\newcommand{\e}{%
	\ensuremath{\mathrm{e}}}

\renewcommand{\imath}{%
	\ensuremath{\mathrm{i}}}

\newcommand{\tsup}[1]{\ensuremath{\mathsf{#1}}}

\newcommand{\ignore}[1]{}

\newcommand{\norm}[1]{\left\lVert#1\right\rVert}

\DeclareMathOperator\erf{erf}

\DeclareMathAlphabet\boldsymbolcal{OMS}{cmsy}{b}{n}

\begin{document}

\title{Statistical Approximations of LOS/NLOS Probability in Urban Environment}

\author{\IEEEauthorblockN{Rimvydas Aleksiejunas}
\IEEEauthorblockA{\textit{Institute of Applied Electrodynamics and Telecommunications, Vilnius University}\\
Vilnius, Lithuania \\
rimvydas.aleksiejunas@ff.vu.lt}
}

\maketitle

\begin{abstract}
Analysis of line-of-sight and non-line-of-sight (LOS/NLOS) visibility conditions is an important aspect of wireless channel modeling. For statistical channel models the Monte Carlo simulations are usually used to generate spatially consistent visibility states based on particular LOS probability. The present works addresses LOS probability approximation problem using a mix of distance-dependent exponential functions for urban areas with high and low building densities. The proposed model divides site coverage area into LOS and NLOS zones approximated by trigonometric series and support vector classification methods. Compared to commonly used generic ITU-R and 3GPP LOS probability models the proposed approximation is more accurate compared to real world LOS distributions. The accuracy of LOS probability model has been tested against visibility predictions obtained from the digital building data over Manhattan and San Francisco city areas.

\end{abstract}


\section{Introduction}

Understanding of radio channel propagation conditions and spatial consistency is important for designing mobile networks especially when moving to  millimeter waves and 5G applications. Many research and standardization efforts are directed by ITU-R \cite{ITU-R-M.2135-09} and 3GPP \cite{3GPP-TR36.873-17} organizations as well as telecommunication companies. The main purpose of radio channel modeling efforts is to build reliable models for radio equipment testing and validation, wireless network planning and compatibility studies. One of the main channel characteristics is the line-of-sight (LOS) probability used in statistical Monte Carlo propagation simulations. LOS probability depends on the distance from base station and this dependence should differ in urban, suburban and rural areas. and as well should take into spatial consistency , \cite{AleksiejunasURSI18}. which is achieved by supplementing statistical propagation models with terrain and building data  by constructing building map-based hybrid prediction models \cite{METIS-D1.4-15}, \cite{SteinbockKyosti16}, \cite{KyostiMapBased17}. Spatial consistency ensures that transition between line-of-sight and non-line-of-sight (NLOS)  is governed by autocorrelation/decorrelation distance. The LOS probability depends on the single parameter - distance from base station. However, usually there is no single homogeneous environments and in real propagation environments both LOS and NLOS conditions are displaced all over the analysis area. In such case dual environment LOS probability model would represent real conditions more accurately.   

Dual environment LOS/NLOS boundary approximation has already been introduced in \cite{AleksiejunasLOS18}, \cite{AleksiejunasURSI18} for Manhattan grid with regular street geometry at the same time maintaining spatial consistency for visibility state distribution. Digital building models of Manhattan city have been used in \cite{AleksiejunasLOS18} to model real-world urban scenarios with trigonometric dual environment boundaries. However, open question still remains on the method of area division into LOS and NLOS parts so that the model remains general enough to be used in radio network simulation studies and at the same time reflecting real propagation conditions.

In the present work, real-world LOS/NLOS visibility statistics is studied based on about 1000 base station locations obtained from US Federal Communications Commission (FCC) antenna tower database for Manhattan and San Francisco cities supplemented by terrain elevation and building heights data. More general statistical distribution characteristics for LOS/NLOS probability is estimated including dual environment with boundary approximations by support vector classification (SVC) method. This allows applications of approximate probability to broader class of urban visibility conditions including mixed high and low building densities.

The structure of the paper is following. In Section~\ref{sec-los-nlos-state}, LOS/NLOS probability models accepted by ITU-R and 3GPP are shortly reviewed including spatial consistency requirements. Then dual environment boundary approximations based on trigonometric series and SVC methods are given in Section~\ref{sec-los-nlos-bound-approx}, followed by statistical estimation of approximation accuracy in comparison to available deterministic visibility model, and finally conclusions are drawn.

\section{LOS/NLOS Visibility State Probability}\label{sec-los-nlos-state}

\subsection{LOS Probability Models}

The most commonly used LOS state probability approximations are based on 3GPP \cite{3GPP-TR36.873-17} and ITU \cite{ITU-R-M.2135-09} 3D channel models representing different propagation scenarios. 3GPP and ITU-R proposed LOS probability models can be approximated by the following distance dependence laws for urban (UMa) and rural (RMa) macrocell areas:
\begin{equation}
p_{\mathrm{LOS}}(d) = \begin{cases}
  \min\left(\frac{d_{1}}{d}, 1 \right) \left(1 - \e^{-\frac{d}{d_{2}}} \right) + \e^{-\frac{d}{d_{2}}}, \; \text{UMa} \\
   d_{3} \e^{-\frac{d}{d_{4}}}, \; \text{RMa}
\end{cases}
\end{equation}
where $d$ is the distance in meters and $d_{i}$, $i=\{1, 2, 3, 4\}$ are empirical data fit coefficients. Similar LOS/NLOS probability models are suggested in other proposals \cite{Wang5GSim14, Ademaj3GPP16} and are discussed in more detail in \cite{AleksiejunasLOS18}. Characteristic feature of such models is the existence of direct LOS visibility region around base station up to distance $d_{1}$ for urban environment.

\subsection{Dual Environment LOS Probability Models}

To better represent real propagation environments which usually contain nonuniform building blocks of varying height, a combined dual environment model which approximates LOS probability over distance using two probability functions was proposed in \cite{AleksiejunasURSI18, AleksiejunasLOS18}. The composite LOS probability is characterized by two separate regions with different LOS probabilities $p^{(1)}_{\mathrm{LOS}}(d)$ and $p^{(2)}_{\mathrm{LOS}}(d)$ can be estimated as
\begin{equation}
p_{\mathrm{LOS}}(d) = p^{(1)}_{\mathrm{LOS}}(d) f(x) + p^{(2)}_{\mathrm{LOS}}(d)\left[1 - f(x)\right], \label{eq-plos-dual}
\end{equation}
where $f(x)$ function indicates density of buildings in zone 1 characterized by LOS probability $p^{(1)}_{\mathrm{LOS}}(d)$ at distance $d$ from the base station, while $x$ denotes the shortest distance to the boundary of zone 1. 

Considering continuous normal distribution $h(x)$ of blocking obstacle heights with mean height $\mu_{h}$ and standard deviation $\sigma_{h}$, the transition function $f(x)$ separating two regions can be expressed by complementary cumulative distribution function as \cite{5GWorkshop16}
\begin{equation}
  f(x) = 1 - \frac{1}{2}\left\{1 - \erf\left[\frac{h(x) - \mu_{h}}{\sqrt{2}\sigma_{h}}\right] \right\}. \label{eq-fx-erf}
\end{equation}
More detailed account on the dual LOS environments are given in \cite{AleksiejunasURSI18}.

There is still an open question about geometry of the boundary dividing site coverage area into different visibility environments. In order to enable usage of such boundary for a channel model, the boundary should be smooth enough and independent of particular local obstacle distribution in order to be used as a representation of a generalized typical environment. In the next section two possible approximations of such LOS/NLOS boundary generalizations are discussed.

\section{LOS/NLOS Boundary Approximations}\label{sec-los-nlos-bound-approx}

Here we present two methods for generalizing LOS/NLOS boundaries obtained from deterministic line-of-sight models which take into account terrain elevation and building heights. After predicting line-of-sight areas from base station antenna locations within given radius the following two approximations are suggested.

\begin{figure}
\begin{subfigure}
  \centering
  \includegraphics[width=0.48\linewidth]{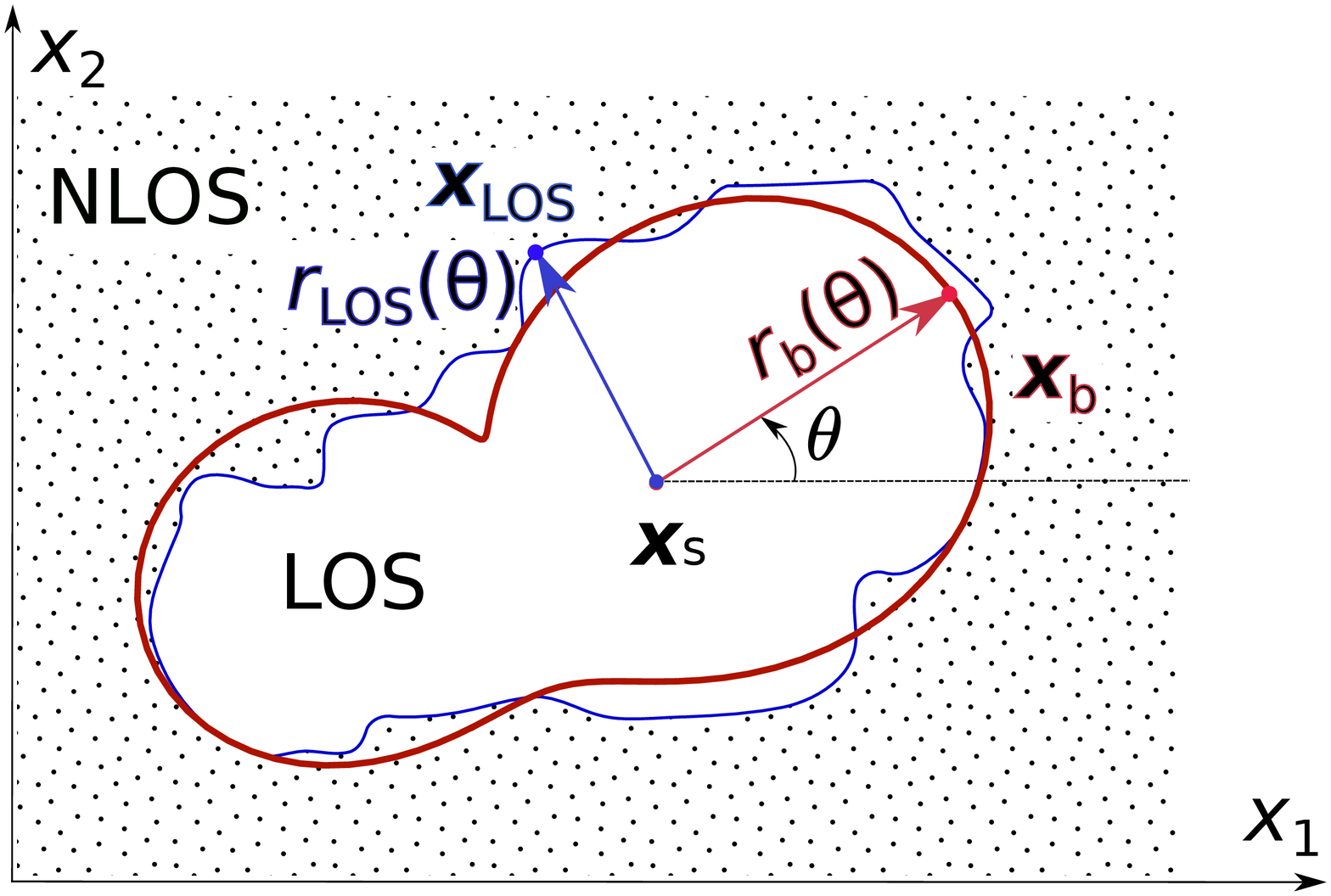}  
  \label{fig-approx-geom-trig}
\end{subfigure}
\begin{subfigure}
  \centering
  \includegraphics[width=0.48\linewidth]{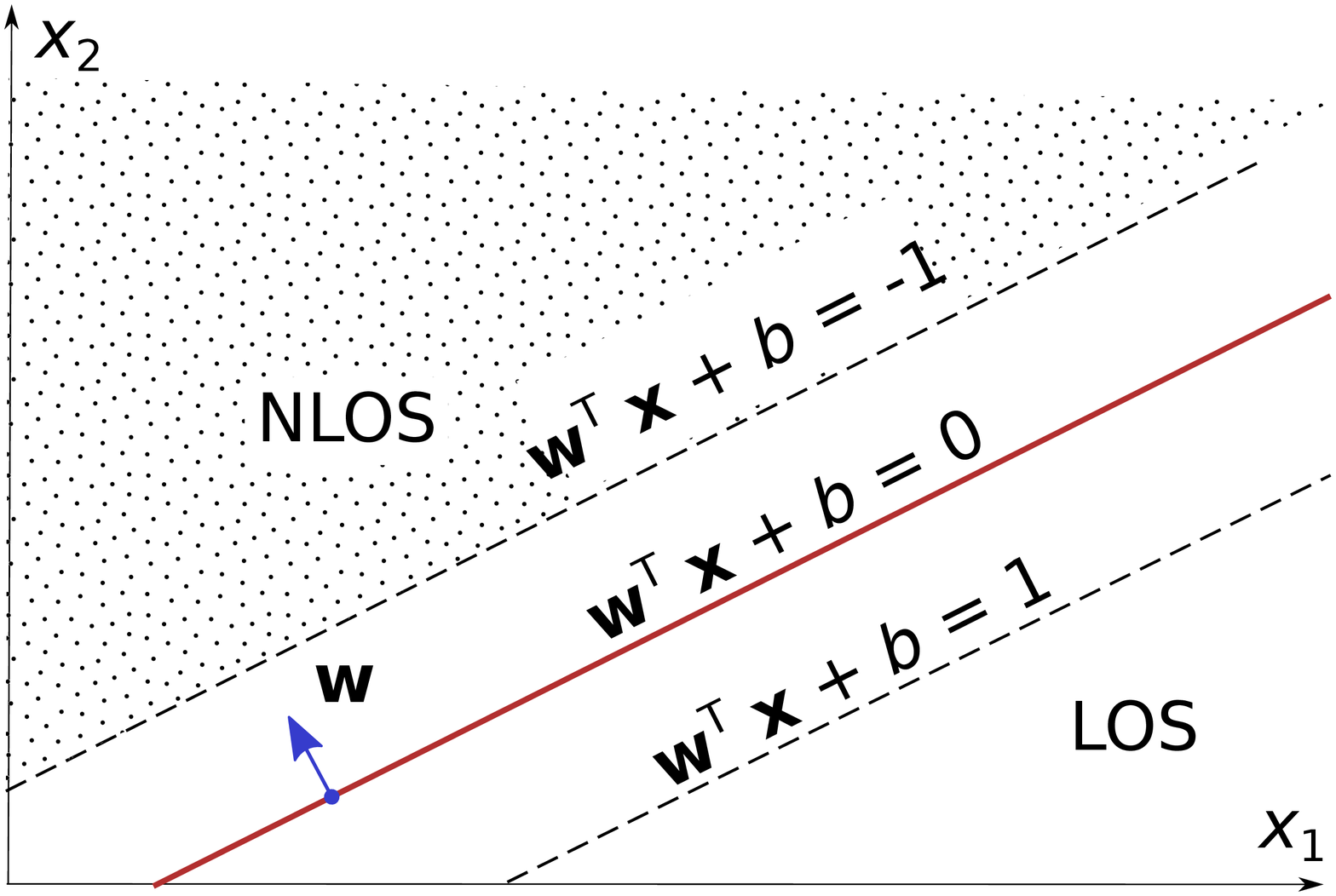}  
  \label{fig-approx-geom-svc}
\end{subfigure}
\caption{LOS/NLOS boundary approximation geometry: trigonometric series (left) and SVC (right).}
\label{fig-approx-geom}
\end{figure}

\subsection{Trigonometric Series Approximation}

We define LOS/NLOS boundary vector $\mathbf{x}_{\mathrm{LOS}} \equiv(x_{1}, x_{2})$ as a parametric equation with radius $r_{\mathrm{LOS}}(\theta)$ over polar angle $\theta \in [0, 2\pi]$ around base station location $\mathbf{x}_{\mathrm{s}}$:
\begin{equation}
  \mathbf{x}_{\mathrm{LOS}} = \mathbf{x}_{\mathrm{s}} + r_{\mathrm{LOS}}(\theta) \left[\cos\theta, \sin\theta \right]^{\tsup{T}},
\end{equation}
the geometry of which is shown in Fig.~\ref{fig-approx-geom} left. To generalize LOS/NLOS boundary we use discrete trigonometric series approximation \cite{ReichelDiscreteTrig91} to the real LOS boundary by smoothed boundary of radius $r_{\mathrm{b}}(\theta)$: 
\begin{equation}
r_{\mathrm{b}}(\theta) = a_{0} + \sum_{n=1}^{N} {a_{n} \cos(n\theta) + b_{n} \sin(n\theta)},
\end{equation}
which minimizes discrete least squares error at $m = 1, \ldots, M$ angular points $\left\{\theta_{m}\right\}$ along the LOS/NLOS boundary $r_{\mathrm{LOS}}(\theta_{m})$ (Fig.~\ref{fig-approx-geom} left):
\begin{equation}
\norm{r_{\mathrm{LOS}}(\theta) - r_{\mathrm{b}}(\theta)} = \left( \sum_{m=1}^{M}{w_{m}^{2} \left| r_{\mathrm{LOS}}(\theta_{m}) - r_{\mathrm{b}}(\theta_{m}) \right|^{2} }\right)^{0.5}.
\end{equation}
For numerical solution of this minimization problem Levenberg-Marquardt algorithm \cite{Levenberg44, Marquardt63} is used as implemented in Python's SciPy library \cite{scipy}. To make LOS zones compact around central base station points the weight coefficients $w_{m}$ are optimized as error penalties $0 \le w_{0} \le 1$ equal for all $m$ points within NLOS zone based on the following condition:
\begin{equation}
  w_{m} = \begin{cases} 
	w_{0},  & r_{\mathrm{LOS}}(\theta_{m}) > r_{\mathrm{b}}(\theta_{m}), \\ 
	0,      & r_{\mathrm{LOS}}(\theta_{m}) \le r_{\mathrm{b}}(\theta_{m}).
    \end{cases}
\end{equation}
Here coefficient $w_{0}$ is kept independent of angular point number $m$ to achieve more general solution with less parameters. LOS/NLOS boundary is described now by optimized trigonometric series coefficients $a_{0}$, $a_{n}$, $b_{n}$, $n=1 ... N$ and optimized value of $w_{0}$. The purpose of $w_{m}=w_{0}$ outside LOS zone is to avoid NLOS blockage centers within central LOS zone in the vicinity of base station. For optimization, fixed number of boundary points $M=100$ and variable length of trigonometric series $N=2,\dots,5$ are used.

\subsection{Support Vector Classification Method}

An alternative method for LOS/NLOS boundary approximation the support vector classification (SVC), namely, $\nu$-SVC \cite{ScholkopfNuSVC00} has been chosen for generating generalized boundaries from the given labeled $y_{i} \in \{-1, +1\} \equiv \{\text{NLOS}, \text{LOS}\} $ dataset $\left\{(\mathbf{x}_{1}, y_{1}), \ldots, (\mathbf{x}_{M}, y_{M})\right\}$ of $M$ points with geometrically estimated visibility conditions based on elevation and building height data. We used rectangular mesh with 2~m step for $M$ points evenly distributed over analysis area around the base station. $\nu$-SVC classification method reduces to minimization problem of objective function for a hyper-plane with the normal $\mathbf{w}$ and bias $b$:
\begin{eqnarray}
&\underset{\mathbf{w}, \xi_{m}, \rho}{\mathrm{min}} & \left(\frac{1}{2} \norm{\mathbf{w}}^{2} - \nu\rho + \frac{1}{M}\sum_{m=1}^{M}\xi_{m}\right) \\
&\mbox{s.t.} & y_{m}\left( \mathbf{w}^{\tsup{T}} \mathbf{x}_{m} \right) + b \ge \rho - \xi_{m}, 	\nonumber\\
& & \xi_{m} \ge 0, \; \rho \ge 0, \; m = 1, \ldots, M, \nonumber 
\end{eqnarray}
with hyper-parameter $\nu \in [0, 1]$ representing upper bound on the misclassified margin error, $\rho$ denoting lower bound on $\norm{\mathbf{w}}$ and $\xi_{m}$ being slack variables. The geometry of hyper-plane in 2D space used to model LOS/NLOS boundaries is shown Fig.~\ref{fig-approx-geom} right. The nonlinear decision function is constructed as a linear combination of support vectors based on Gaussian kernel $k\left(\mathbf{x}, \mathbf{x}_{m}\right) = \exp\left(-\gamma \norm{\mathbf{x} - \mathbf{x}_{m}}^{2} \right)$. The $\gamma$ parameter controls the smoothness of approximated boundary which can be expressed via spatial deviation $\sigma$ as $\gamma = 1/2\sigma^{2}$. Taking into account that the decorrelation distance due to shadowing may reach up to 50~m \cite{5GWorkshop16}, the generalized boundary has to be defined by lower spatial variation, therefore for simulations  $\gamma=10^{-4}$~m\textsuperscript{-2} is chosen corresponding to spatial deviation $\sigma=71$~m. To improve performance of SVC classification the ensemble learning -- bootstrap aggregating \cite{LouppeEnsembles12} is used as implemented in Python's scikit-learn library \cite{PedregosaScikitLearn11}. The number of ensemble estimators during SVC optimization has been varied between 10 and 40.

\subsection{Statistical Results of LOS Probability Approximations}

About 1000 base station locations from Manhattan and San Francisco cities have been used to generate deterministic line-of-sight coverages around base stations within 500~m radius taking into account base station antenna heights, terrain elevation data and building heights. Antenna tower data has been obtained from US FCC Antenna Structure Registration database \cite{USFCC_ASR_20}. Digital elevation model (DEM) of 1/3 arc-second resolution available from US Geological Survey (USGS) \cite{USGS_3DEP_17} has been used for terrain modeling. Building heights are extracted from building footprint datasets provided by open data initiatives of New York \cite{Bld_NYC_19} and San Francisco \cite{Bld_SF_19} cities. For visibility calculations DEM raster has been resampled to 10~m resolution and combined with height information extracted from vector-type building footprints. The resolution of combined surface raster has been set to 2~m which is the final resolution of all visibility predictions presented here.

\begin{figure}
\centering
\includegraphics[width=0.48\textwidth]{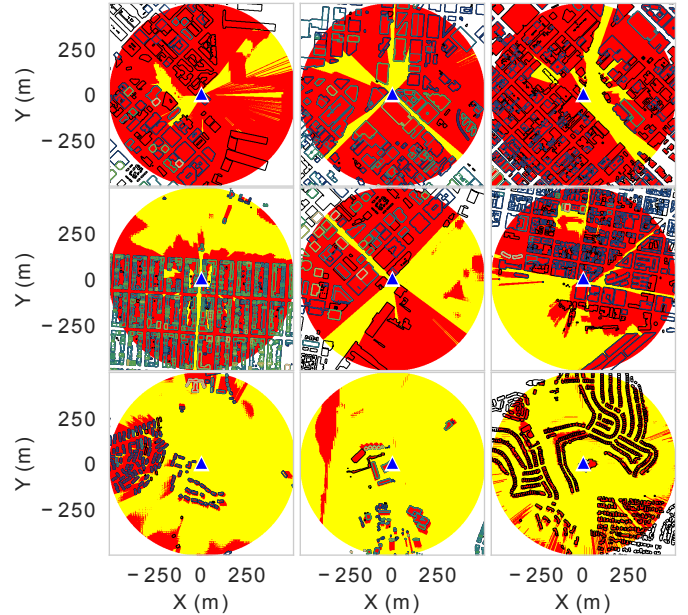}
\caption{Typical urban cell sites in San Francisco city with different percentages of line-of-sight visibility: 10\% top row, 30\% middle row and 50\% bottom row. Yellow color indicates visible (LOS) areas, red color -- NLOS areas. Contours represent building footprints and rectangles denote base station locations.}
\label{fig-blos-3x3}
\end{figure}

A typical selection of LOS estimation results for different visibility conditions is shown in Fig.~\ref{fig-blos-3x3}. Here the top row shows sites with about 10\% directly visible LOS locations within site coverage, the middle row corresponds to LOS/NLOS fraction of 30\% and the bottom row contains sites with 50\% or more open locations.   

\begin{figure}
\centering
\includegraphics[width=0.48\textwidth]{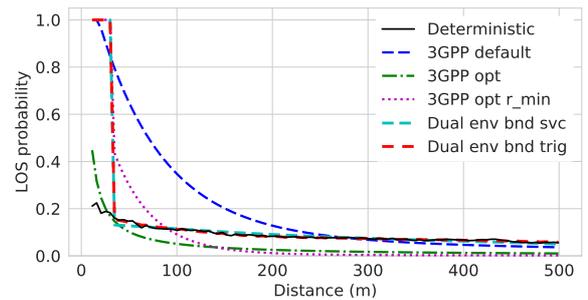}
\caption{LOS probability approximation by different methods for urban area with 30\% LOS visibility. 2D plot of this area is shown in the middle row, middle column of Fig.~\ref{fig-blos-3x3}.}
\label{fig-los-prob-approx}
\end{figure}

Then for each of base station locations the LOS probability $p_{\mathrm{LOS}}(d)$ dependency on distance $d$ is estimated over whole cell area and approximated by several methods: (i) by single exponential with default parameter values as in 3GPP (1); (ii) with single exponential but fitted parameters $d_{i}$; (iii) with fitted 3GPP exponential parameters and having optimized minimum LOS distance $d_{1}$; and (iv) using dual environment boundary approximations by trigonometric series and SVC method. The difference between deterministic LOS probability and estimated by various model approximations is shown in Fig.~\ref{fig-los-prob-approx} for a single site which 2D coverage is represented in the center of Fig.~\ref{fig-blos-3x3} with 30\% direct LOS visibility. From the visual comparison of these results, all the approximations start at LOS probability equal to $p_{\mathrm{LOS}} = 1.0$ at the base station location $d = 0$ except the case of all 3GPP parameters being optimized. Although this approximation follows deterministic LOS probability most closely it lacks 3GPP requirement to have immediate line-of-sight area within base station's close vicinity.

\begin{figure}
\centering
\includegraphics[width=0.5\textwidth]{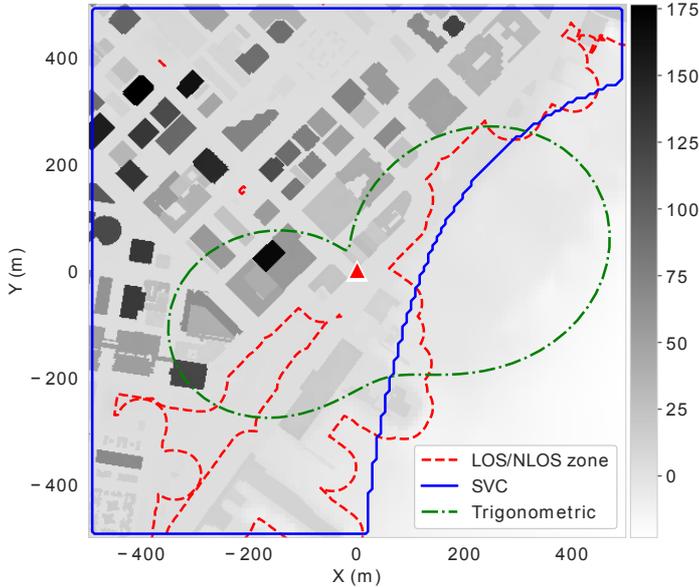}
\caption{LOS/NLOS boundary approximations by trigonometric series and SVC method. Dashed red line indicates urban build-up area with 40~m buffer zone. In the background, surface height consisting of terrain elevation and building heights is shown proportionally in gray color.}
\label{fig-los-bounds-approx}
\end{figure}

For this specific site represented by Fig.~\ref{fig-los-prob-approx} LOS probability approximations, a 2D plot of LOS/NLOS boundary approximations by trigonometric series and SVC classification model is depicted in Fig.~\ref{fig-los-bounds-approx}. For this specific case least squares trigonometric minimization problem resulted in series order $N = 2$ and weights coefficient $w_{0} = 0.5$, while SVC optimization resulted in $\nu = 0.5$ and number of ensemble estimators 20. The approximation RMSE errors for trigonometric series and SVC are, respectively, 0.18 and 0.17. These errors belong to the worst end of LOS/NLOS boundary approximation statistics where mixed LOS/NLOS conditions predominate. The staircase look of SVC approximation is due to rasterized sampling of the whole analysis area by 2~m size pixels. For comparative purposes, in the same figure, red contour denotes 40~m buffered zone indicating deterministic LOS/NLOS boundary which is over-complex and too specific for particular urban environment in order to be used for generalized LOS probability models.  

\begin{figure}
\centering
\includegraphics[width=0.5\textwidth]{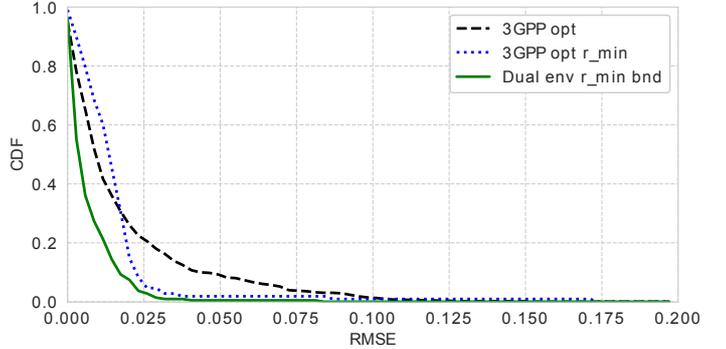}
\caption{Total CDF statistics of LOS probability approximation RMSE errors for all line-of-sight visibility cases used in the analyzed urban site dataset: 3GPP with optimized coefficients (dashed line), 3GPP with optimized coefficients and minimum LOS distance $d_{1}$ (dotted line) and dual environment LOS model with minimum radius (solid line). Here dual environment CDF includes both trigonometric series and SVC approximations.}
\label{fig-cdf-rsme}
\end{figure}

Approximation errors for each method depend on the visibility conditions -- the fraction of directly visible locations, but the total cumulative distribution function (CDF) of RMSE errors for all base stations clearly shows advantage of dual environment approximations as shown in Fig.~\ref{fig-cdf-rsme}. The mean RMSE for 3GPP optimized exponential model is 0.020, or 0.018 if minimum LOS distance is included, and 0.010 for dual environment models approximated by trigonometric series or SVC method, while in all cases lower than the previous example. 

\begin{figure}
\centering
\includegraphics[width=0.5\textwidth]{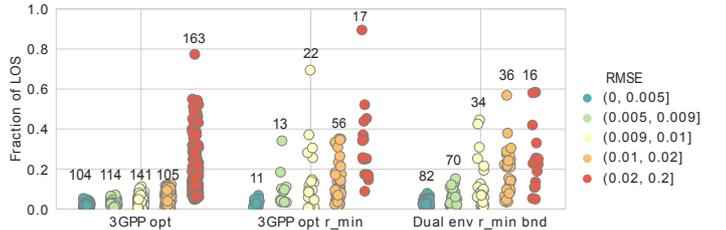}
\caption{RMSE distribution over all analyzed base station locations for the best LOS approximations by different methods with respect to different LOS visibility conditions. The numbers at the top of each group indicate total number of points in the group.}
\label{fig-los-frac}
\end{figure}

More thorough picture about LOS probability approximation accuracy can be composed by spreading RMSE errors for different methods over a range of different LOS visibility conditions. In this case the best approximation results with lowest RMSE are grouped along the extra axis of direct LOS visibility as shown in Fig.~\ref{fig-los-frac}. Here for each base station the best accurate approximation is selected and resultant RMSE error distribution is divided into five quantiles which are stretched along LOS visibility axis. While the traditional single exponential 3GPP model works well (RMSE being less than 0.02) in mostly NLOS cases with LOS fraction below 0.17, the rest visibility range results in RMSE errors between 0.02 and 0.2. The other two approximations, 3GPP optimized and especially dual environment model have significant proportion of mid-range LOS fraction between 0.2 and 0.6 covered by approximation RMSE errors below 0.02.

Although most methods have tendency to work best with dense urban NLOS conditions, the dual environment approximations tend to be more suitable for intermediate visibility conditions, where large portions of up to 40-60\% are attributed to open areas.

\section{Conclusions}

The statistical results of LOS probability gathered for urban areas in San Francisco and Manhattan cities support possibility of using dual environment model approximations with boundaries based on trigonometric series or SVC classification. Such approximations are especially advantageous at higher percentages of directly visible areas within base station coverage. These LOS conditions indicate nonuniform LOS environments where high and low density urban regions are located. Dual environment approximations could be used in combination to single exponential LOS probability models commonly used to homogeneous environments such as urban, suburban or rural areas. In this case more complexity of dual environment model results in higher accuracy simulation model at the same time maintaining generality and spatial consistency required by wireless channel models.

\IEEEtriggeratref{10}
\bibliographystyle{IEEEtran}

\end{document}